\documentstyle[12pt]{article}
\begin{document}

\def\beq{\begin{equation}}
\def\eeq{\end{equation}}
\def\bce{\begin{center}}
\def\ece{\end{center}}
\def\bea{\begin{eqnarray}}
\def\eea{\end{eqnarray}}
\def\ben{\begin{enumerate}}
\def\een{\end{enumerate}}
\def\ul{\underline}
\def\ni{\noindent}
\def\nn{\nonumber}
\def\bs{\bigskip}
\def\ms{\medskip}
\def\wt{\widetilde}
\def\wh{\widehat}
\def\brr{\begin{array}}
\def\err{\end{array}}
\def\Tr{\mbox{Tr}\ }

\hfill HUPD-9413


\vspace*{3mm}

\begin{center}

{\LARGE \bf
A four-dimensional theory for quantum gravity with conformal and
nonconformal explicit solutions }

\vspace{4mm}

\medskip

 {\sc E. Elizalde} \\
 Center for Advanced Study, CEAB, CSIC, Cam\'{\i} de Santa B\`arbara, 17300
Blanes \\ and Department ECM and IFAE, Faculty of Physics,
University of  Barcelona \\ Diagonal 647, 08028 Barcelona, Catalonia,
Spain\\
  and Department of Physics,
 Hiroshima University,  Higashi-Hiroshima 724,
Japan \footnote{Address june-september 1994. E-mail:
 eli@zeta.ecm.ub.es}\\

\vskip 5mm

 {\sc A.G. Jacksenaev} \\
Tomsk Pedagogical Institute, 634041 Tomsk, Russia \\

\vskip 5mm

{\sc S.D. Odintsov} \footnote{E-mail: odintsov@ecm.ub.es}\\
Department ECM, Faculty of Physics,
University of  Barcelona, \\  Diagonal 647, 08028 Barcelona, Catalonia,
Spain \\
and Tomsk Pedagogical Institute, 634041 Tomsk, Russia, \\

\vskip 5mm

and
{\sc I.L. Shapiro} \footnote{Present address: Departamento de
F\'{\i}sica
Te\'orica, Universidad de Zaragoza, 50009 Zaragoza, Spain. E-mail:
shapiro@dftuz.unizar.es}\\
Tomsk Pedagogical Institute, 634041 Tomsk, Russia, \\
and Department of Physics, Hiroshima University, Higashi-Hiroshima
724, Japan

\vspace{5mm}

{\bf Abstract}

\end{center}

The most general version of a renormalizable $d=4$ theory corresponding
to a dimensionless
higher-derivative scalar field model in curved spacetime is explored.
The classical action of the
theory contains $12$ independent functions, which are the generalized
 coupling
constants of the theory. We calculate the one-loop beta functions and then
consider the conditions for finiteness. The set of exact
solutions of power type is proven to consist of precisely three
 conformal and three
nonconformal solutions, given by remarkably simple (albeit nontrivial)
functions that we obtain explicitly. The finiteness
of the conformal theory indicates the absence of a conformal anomaly in
the finite sector.
The stability of the finite solutions is investigated and the
 possibility of renormalization group
 flows is discussed as well as several physical applications.

\vspace{4mm}

\newpage

\section{Introduction}

The considerable achievements that have been obtained  in the field of
two-dimensional quantum gravity have
inspired different attempts to use it as a pattern for the
construction of the more
realistic theory of quantum gravity in four dimensions. Unfortunately
the direct analogies of the two cases
do not work here, for rather evident reasons. First of all, the quantum
metric in $d=4$ has more degrees of freedom, which include the
physical
degrees of freedom of spin two, what is quite different from the $d=2$
case. Second,
the Feynman integrals in $d=4$ have worst convergence properties as
compared with the $d=2$ case,
from what follows that higher-derivative terms have to be included in
order to  ensure
renormalizability. An example of this sort is given by quantum
$R^2$-gravity (for a review and a list of references see \cite{3}),
which is multiplicatively renormalizable \cite{9} (not so is Einstein's
gravity) and also asymptotically free.
However the presence of higher derivatives leads to
the problem of massive spin-two ghosts, which violate the unitarity of the
$S$-matrix. It has been conjectured, nevertheless, that the
problem of non-unitarity in R$^2$-gravity might perhaps be solved in a
non-perturbative approach.

The alternative approach is based on the assumption that gravity is the
induced interaction and the equations for the gravitational field arise
as effective ones in some more general theory, as the theory of
(super)strings \cite{1}.
It is also interesting to notice that higher-derivative gravitational
theories (like string-inspired models) often admit singularity-free
solutions (for a recent discussion and a list of references, see
\cite{4,5}).
In string theory,  higher-derivative actions also arise in quite a
natural way. For instance, if one
wants to study the massive higher-spin modes of the theory one has to
modify
the standard $\sigma$-model action by adding to it an infinite number of
terms, which contain all possible derivatives.
On the other hand, the effective action of gravity, which follows from
string theory, contains higher-derivative terms, and the higher powers
in
derivatives correspond to the next order of string perturbation theory.
One can expect that the unitarity of the theory will be restored when
all the
excitations are taken into account. Therefore, it is quite natural to
consider
fourth-order gravity as some kind of effective theory, which is valid as
an approximation to a more fundamental theory, still unknown.

String-inspired models of gravity contain, at least, two independent
fields, which are the
metric and  the scalar dilaton field. Hence, the aforementioned
effective theory
has to depend on the dilaton field as well.
 The more general action (\ref{1x})
for a renormalizable theory of this type has been recently formulated in
\cite{6}. Since this model is rather complicated, even the one-loop
calculations
are very tedious. At the same time it is possible to make quite a
considerable
simplification: since both the metric and the dilaton are dimensionless,
higher-derivative fields, the structure of divergences is essentially
the same even
if the metric is taken as a purely classical background. Indeed, the
renormalization
constants are different, if compared with the complete theory, but their
general structures have to be similar.

 Let us recall that the theory of a
quantum dilaton field has been recently proposed for the description of
infrared
quantum gravity \cite{antmot} (see also \cite{28} and \cite{27}).
Furthermore it has turned out that
the quantum dilaton theory enables one to estimate the back reaction of
the vacuum
to the matter fields \cite{shcog}. It is very remarkable that the effect
of the quantum
dilaton is qualitatively the same as the effect of the quantum metric,
evaluated earlier in \cite{buchodsh}.
In a previous article \cite{6} we have considered the
one loop renormalization and asymptotic behaviour of the special constrained
version of the dilaton theory. In fact the action of
this special model is
the direct extension of the action for induced gravity
\cite{rei,frts,first,antmot}.
In particular,
we have found that this constrained model has induced gravity
as the renormalization-group fixed-point, and that it also exhibits
asymptotical conformal invariance.

The present paper is devoted to the study of the quantum properties of
the most general
higher-derivative scalar theory in curved spacetime.
The paper is organized as follows. Section 2 contains a brief
description of
the model. In section 3 we calculate the one-loop divergences with
the use of the
standard Schwinger-DeWitt technique, which is
modified a little, in accordance to the needs
of our higher-derivative dilaton theory.
Sections 4 and 5 are devoted to the search for all the one-loop finite
solutions (of a specific power-like type) of the renormalization
group (RG) equations. First of all, we consider the conformal version of
dilaton gravity
 (this model
is an extension of the one formulated in \cite{17,rei})
and thus construct three different examples of anomaly-free dilaton
models.
Then the more general nonconformal version is explored. In section 6 we
present some analysis of the asymptotic behaviour of the theory,
together with a number of
 mathematical tools which are useful in this field. Section 7 contains
the discussion of our results, including the possible role of the effects
 of the quantum metric.

\medskip

\section{Description of the model}

We start with
an action of $\sigma$-model type which is renormalizable in a
generalized sense. A basic assumption will be that the scalar
$\varphi$ be dimensionless
in four-dimensional curved spacetime, namely that
$[\varphi ]=0$. We will also admit that there is just
 one fundamental dimensional constant, which
has dimension of mass squared. The only field, aside
from the scalar, which will be present in the theory is the
gravitational field $g_{\mu\nu}$.

Then, dimensional
considerations lead us to the following general action of sigma-model
type
$$
S= \int d^4x \sqrt{-g} \{ b_1 (\varphi ) ( \Box
\varphi )^2 +  b_2 (\varphi ) \left( \nabla_\mu \varphi
\right) \left( \nabla^\mu \varphi \right)
 \Box \varphi +  b_3 (\varphi ) [ ( \nabla_\mu \varphi
)( \nabla^\mu \varphi ) ]^2
 $$
$$
 +  b_4 (\varphi ) ( \nabla_\mu \varphi ) (
\nabla^\mu \varphi )+  b_5 (\varphi ) +  c_1 (\varphi ) R
( \nabla_\mu \varphi ) ( \nabla^\mu \varphi
) +  c_2 (\varphi ) R^{\mu\nu} ( \nabla_\mu \varphi
) ( \nabla_\nu \varphi )
 $$
\beq
+ c_3 (\varphi ) R \Box \varphi + a_1 (\varphi )
R^2_{\mu\nu\alpha\beta }  + a_2 (\varphi ) R^2_{\mu\nu} + a_3
(\varphi ) R^2 + a_4 (\varphi ) R \}
+ (\mbox{s.t.}),       \label{1x}
\eeq
where $s.t.$ means `surface terms'.
All generalized coupling constants are
dimensionless, except for $b_4$, $b_5$ and $a_4$, for which we
have:  $[b_4 (\varphi )]=2$,  $[b_5 (\varphi )]=4$,  $[a_4
(\varphi )]=2$. All other possible terms that can appear in
dimension 4 in the above model can be obtained from (\ref{1x})
by simple integration by parts, and thus differ from these structures
of the above action by some surface terms (s.t.) only. One can
easily verify the following reduction formulas
$$
c_4(\nabla^\mu R)(\nabla_\mu\varphi)=-c'_4R(\nabla_\mu\varphi)^2-c_4R
(\Box\varphi) + (s.t.)
$$
$$
c_5(\Box R)=c''_5R(\nabla_\mu\varphi)^2+c'_5R(\Box\varphi) + (s.t.)
$$
$$
c_6R_{\mu\nu}(\nabla^\mu\nabla^\nu\varphi)=-c'_6R_{\mu\nu}
(\nabla^\mu\varphi)(\nabla^\nu\varphi)+{1\over 2}
c'_6R(\nabla_\mu\varphi)^2+{1\over 2}c_6R(\Box\varphi)  + (s.t.)
$$
$$
b_6(\nabla^\nu\varphi)(\Box\nabla_\nu\varphi)=-b'_6(\nabla_\mu\varphi)^2
(\Box\varphi)-b_6(\Box\varphi)^2+b_6R_{\mu\nu}(\nabla^\mu\varphi)
(\nabla^\nu\varphi) + (s.t.)
$$
$$
b_7(\nabla_\nu\nabla_\mu\varphi)^2={1\over 2}b''_7(\nabla_\mu\varphi)^4+
{3\over 2}b'_7(\nabla_\mu\varphi)^2(\Box\varphi)+b_7(\Box\varphi)^2-b_7
R_{\mu\nu}(\nabla^\mu\varphi)(\nabla^\nu\varphi)  + (s.t.)   \label{2x}
$$
$$
b_8(\nabla_\nu\varphi)(\nabla_\mu\varphi)(\nabla^\nu\nabla^\mu\varphi)=
(-{1\over 2})[b'_8(\nabla_\mu\varphi)^4+b_8(\nabla_\mu\varphi)^2(\Box\varphi)]
 + (s.t.)
$$
$$
b_9(\nabla^\nu\Box\nabla_\nu\varphi)=b''_9(\nabla_\mu\varphi)^2(\Box\varphi)+
b'_9(\Box\varphi)^2-b'_9R_{\mu\nu}(\nabla^\mu\varphi)(\nabla^\nu\varphi)
 + (s.t.)
$$
$$
b_{10}(\Box^2\varphi)=b''_{10}(\nabla_\mu\varphi)^2(\Box\varphi)+
b'_{10}(\Box\varphi)^2 + (s.t.)
$$
\[
b_{11}(\nabla^\nu\varphi)(\nabla_\nu\Box\varphi)=-b'_{11}(\nabla_\mu\varphi)^2
(\Box\varphi)-b_{11}(\Box\varphi)^2 + (s.t.)
\]
Here $c_{4,5,6} = c_{4,5,6}(\varphi),b_{6,...,11} = b_{6,...,11}(\varphi)$
are some (arbitrary) functions.
We shall extensively use these formulas below. Notice that, for constant
$\varphi$, this theory represents at the classical level the standard
R$^2$ gravity.

Theory (\ref{1x}) is renormalizable in a generalized sense, i.e.,
assuming that the form of the scalar functions $b_1(\varphi), \ldots,
a_4(\varphi)$ is allowed to change under renormalization. As we see,
also some terms corresponding to a new type of the non-minimal
scalar-gravity interaction appear, with the generalized non-minimal
couplings $c_1(\varphi), c_2(\varphi)$ and $c_3(\varphi)$.

It is interesting to notice that, at the classical level and for some
particular choices of the generalized couplings, the
 action (\ref{1x}) may be viewed, in principle, as a superstring
theory effective action ---the only background fields being the
gravitational
field and the dilaton, see \cite{1}. It has been known for some time
that string-inspired effective theories with a massless dilaton lead
to  interesting physical consequences, as a cosmological
variation of the fine structure constant and of the gauge couplings
\cite{1}, a violation of the weak equivalence principle \cite{7}, etc.
It could seem that all these effects are in conflict with existing
experimental data.
However, some indications have been given \cite{8} that
non-perturbative loop effects might open a window for the existence of
the dilaton, beeing perfectly compatible with the known experimental
data.
This gives good reasons for the study of  higher-derivative
generalizations of theories of the Brans-Dicke type \cite{2} and, in
particular, of their quantum structure.

\medskip

\section{Calculation of the counterterms}

In this section we shall present the details of the calculation of the
one-loop counterterms of the theory for the dilaton
in an external gravitational
field.
For the purpose of calculation of the divergences we will
apply the background field method and the Schwinger-De Witt technique.
The features of higher-derivative theories do not allow for
the use of the last
method in its original form. At the same time,  a few examples of
calculations in higher-derivative  gravity theory are known
\cite{10}--\cite{15} (see also \cite{3} for a
review and more complete list of references) which possess a more
complicated
structure than (\ref{1x}), because of the extra diffeomorphism symmetry.
Let us start with the usual splitting of the field
into background $\varphi$
and quantum $\sigma$ parts, according to
\beq
\varphi \rightarrow \varphi' = \varphi + \sigma.     \label{z1}
\eeq
The one-loop
effective action is given by the standard general expression
\beq
\Gamma = {i \over 2} \Tr \ln {H},               \label{z2}
\eeq
where $H$ is the bilinear form of the action (\ref{1x}).
Substituting
(\ref{z1}) into (\ref{1x}), and taking into account the bilinear part of the
action only, after
making the necessary integrations by parts (the surface terms give no
contribution to $\Gamma$), we obtain the following self-adjoint bilinear form:
\beq
H=2b_1 ( \Box^2+L^{\alpha\beta\gamma}\nabla_\alpha\nabla_\beta
\nabla_\gamma+V^{\alpha\beta}\nabla_\alpha\nabla_\beta+
N^\alpha\nabla_\alpha+U ),
\label{z3} \eeq
where the $L^{\alpha\beta\gamma}$ have the specially simple
structure $$
L^{\alpha\beta\gamma}\nabla_\alpha\nabla_\beta\nabla_\gamma={1 \over{2b_1}}
[4b'_1 (\nabla_\mu\varphi)\nabla^\mu\Box] = L^\lambda
\nabla_\lambda\Box.
$$
The quantities $ V^{\alpha\beta}, N^\alpha$ and $U$ are defined
according to
$$
V^{\alpha\beta}\nabla_\alpha\nabla_\beta={1 \over{b_1}}\{[(c'_3-c_1)R+(3b'_1-
2b_2)(\Box\varphi)+(b''_1-2b_3)(\nabla_\mu\varphi)^2-b_4m^2]\Box $$
$$
+ [-c_2R_{\mu\nu}+(2b'_2-4b_3)(\nabla_\mu\varphi)(\nabla_\nu\varphi)+2b_2
(\nabla_\mu\nabla_\nu\varphi)]\nabla^\mu\nabla^\nu\}$$
$$
N^\alpha\nabla_\alpha={1 \over{b_1}}\{(c''_3-c'_1)R(\nabla_\mu\varphi)
\nabla^\mu+(c'_3-{1\over 2}c_2-c_1)(\nabla_\mu R)\nabla^\mu+(2b_2-c'_2)
R_{\mu\nu}(\nabla^\mu\varphi)\nabla^\nu $$
$$
+2(b''_1-2b_3)(\Box\varphi)(\nabla_\mu\varphi)\nabla^\mu+2(b''_2-3b'_3)
(\nabla_\nu\varphi)^2(\nabla_\mu\varphi)\nabla^\mu+4(b'_2-2b_3)
(\nabla_\nu\varphi)(\nabla^\nu\nabla^\mu\varphi)\nabla_\mu $$
$$
+ 2b'_1(\nabla^\mu\Box\varphi)\nabla_\mu-b'_4m^2(\nabla_\mu\varphi)
\nabla^\mu\},
$$
$$
U={1 \over{b_1}} \{(c''_3-c'_1)R(\Box\varphi)+(c''_3-{1\over 2}c'_2-c'_1)
(\nabla_\mu R)(\nabla^\mu\varphi)+({1\over 2}c'''_3-{1\over 2}c''_1)R
(\nabla_\mu\varphi)^2
$$
$$
+(b'_2-{1\over 2}c''_2)
R_{\mu\nu}(\nabla^\mu\varphi)(\nabla^\nu\varphi)+ (b'''_1-2b'_3)
(\nabla_\mu\varphi)^2(\Box\varphi)+({3\over 2}b''_1-b'_2)(\Box\varphi)^2+$$ $$
+({1\over 2}b'''_2-{3\over 2}b''_3)(\nabla_\mu\varphi)^4+(2b''_2-4b'_3)
(\nabla_\mu\varphi)(\nabla_\nu\varphi)(\nabla^\mu\nabla^\nu\varphi)-$$  $$-
c'_2R_{\mu\nu}(\nabla^\mu\nabla^\nu\varphi)+b'_1(\Box^2\varphi)
$$
$$
+2b''_1 (\nabla_\mu\varphi)(\nabla^\mu\Box\varphi)+b'_2(\nabla_\mu\nabla_\nu
\varphi)^2+{1\over 2}c'_3(\Box R)-{1\over
2}b''_4m^2(\nabla_\mu\varphi)^2 $$
\beq
-b'_4m^2(\Box\varphi)+{1\over 2}a''_1R^2_{\alpha\beta\gamma\tau}
+{1\over 2}a''_2R^2_{\alpha\beta}+{1\over 2}a''_3R^2+
{1\over 2}a''_4m^2R+
{1\over 2}b''_5m^4\}.
\label{z4}
\eeq

The next problem is to separate the divergent part of the trace (\ref{z2}).
First of all, let us note that (\ref{z3}) is just a particular case of
the general fourth-order
operator which has been considered in \cite{16}. However, direct use of the
general results in \cite{16} leads to very cumbersome calculations and
we use a
different procedure, already employed in \cite{11}. Let us rewrite the
trace (\ref{z4}) under the form
\beq
\Tr \ln H =
\Tr\ln (2b_1) + \Tr\ln (\Box^2+L^{\mu}\nabla_\mu \Box
+V^{\mu\nu}\nabla_\mu\nabla_\nu +
N^\mu\nabla_\mu +U), \label{z5}
\eeq
and notice that the first term does not give contribution to
the divergences.
Let us explore  the second term. From standard considerations based on
power counting and covariance, it
 follows that the possible divergences have the form
$$
\Tr\ln(\Box^2+L^{\mu}\nabla_\mu \Box
+V^{\mu\nu}\nabla_\mu\nabla_\nu +
N^\mu\nabla_\mu +U)|_{div}
$$ $$
=\Tr\{k_1U+
k_2L^\lambda N_\lambda +
k_3L^\lambda\nabla_\lambda V-
k_4L^\lambda\nabla^\tau V_{\lambda\tau} $$
$$
-k_5VL_\lambda L^\lambda-
k_6V_{\lambda\tau}L^\lambda L^\tau+
k_7\nabla_\lambda L_\tau\nabla^\lambda L^\tau
+k_8\nabla_\lambda L_\tau\nabla^\tau L^\lambda
$$
$$ +k_9L^\tau L^\lambda\nabla_\tau L_\lambda+
k_{10}L_\lambda L_\tau L^\lambda L^\tau
+ k_{11}R^2_{\mu\nu}+
k_{12}R^2
$$
\beq
+k_{13}RV+
k_{14}R_{\mu\nu}V^{\mu\nu}+
k_{15}V^2+
k_{16}V_{\mu\nu}V^{\mu\nu}\} +
 (\mbox{s.t.}), \label{z6} \eeq
where $k_{1...16}$ are some (unknown) divergent coefficients.

The questions is now to find their explicit values in the
one-loop approximation.
It is easy to classify the terms in (\ref{z6}) into several groups.
The first group is formed by the structures the structures with
numerical factors $k_{1,11,...,16}$ ---those are the
ones which do not depend on $L^{\mu}$.
The divergences of this type are just the same as for the operator
\beq
\Box^2 +V^{\alpha\beta}\nabla_\alpha\nabla_\beta+
N^\alpha\nabla_\alpha+U,
\label{z7} \eeq
and we can use the well-known values from \cite{11}.
To the second group belong the structures with $k_{7...10}$.
Here we will use the following method. Since these structures
do not contain $V, N$ and $U$, it is clear that
$k_{7...10}$ will be just the same as for
(\ref{z6}) with $V = N = U = 0$. Hence we can simply put $V = N = U =
0$. Then, taking into account that $
L^{\alpha\beta\gamma}\nabla_\alpha\nabla_\beta
\nabla_\gamma = L^\lambda\nabla_\lambda\Box$, we can write
\beq
\Tr\ln (\Box^2+L^{\alpha}\nabla_\alpha\Box)=
\Tr\ln (\Box) +
\Tr\ln (\Box+L^{\alpha}\nabla_\alpha). \label{z8}
\eeq
The first term gives contribution  to the $k_{11,12}$ only, which we
have
already taken into account. The second term has a standard structure,
 and its contribution has a well-known form (see, for example, \cite{3}).

The third group is just the mixed sector with coefficients $k_{2...6}$.
Here we use the following method \cite{14}. Performing the
transformation $$
\Tr\ln (\Box^2+L^{\alpha}\nabla_\alpha\Box
+V^{\alpha\beta}\nabla_\alpha\nabla_\beta+ N^\alpha\nabla_\alpha+U)
$$
\beq
=\Tr\ln (1+L^{\alpha}\nabla_\alpha\Box
^{-1}+V^{\alpha\beta}\nabla_\alpha\nabla_\beta
\Box^{-2}+ N^\alpha\nabla_\alpha\Box^{-2}+U\Box^{-2})
+ \Tr\ln (\Box^2),      \label{z9}
\eeq
we can easily find that the second term contributes only to $k_{11,12}$.
Then we can expand the
logarithm in the first term into a power series (see \cite{3} for
details)
and use the universal traces of \cite{16}. After a little algebra,
we obtain the final result in the form:
$$
\Tr\ln H=\frac{2i}{\varepsilon}\Tr\{-U+{1\over 4}L^\lambda
N_\lambda+{1\over 6}
L^\lambda\nabla_\lambda V-{1\over 6}L^\lambda\nabla^\tau V_{\lambda\tau}$$
$$ -{1\over 24}VL_\lambda L^\lambda-{1\over 12}V_{\lambda\tau}L^\lambda
L^\tau+{1\over 2}P^2+{1\over 12}S_{\mu\nu}S^{\mu\nu}+{1\over 30}
R^2_{\mu\nu}
$$ $$+{1\over 60}R^2+{1\over 12}RV-{1\over 6}R_{\mu\nu}V^{\mu\nu}+
{1\over 48}V^2+{1\over 24}V_{\mu\nu}V^{\mu\nu}\}, $$
where
$$
P={1\over 6}R-{1\over 2}\nabla_\lambda L^\lambda-{1\over 4}L_\lambda
L^\lambda\;,\;V=V^\mu_\mu\;,$$ \beq
S_{\mu\nu}={1\over 2}(\nabla_\nu L_\mu-\nabla_\mu L_\nu)+{1\over 4}
(L_\nu L_\mu-L_\mu L_\nu).  \label{z10}
\eeq

Finally, substituting (\ref{z4}) into (\ref{z10}) and after a very
tedious
algebra which uses the reduction formulas (\ref{2x}), we
arrive at the following result:
$$
\Gamma^{(1-loop)}_{div}= -{2\over\varepsilon}
\int d^4x\sqrt{-g}[A_1R^2_{\alpha\beta\gamma\tau}+A_2R^2_{\alpha\beta}+
A_3R^2+A_4Rm^2
$$
$$
+ C_1R(\nabla_\mu\varphi)^2+C_2R_{\mu\nu}
(\nabla^\mu\varphi)(\nabla^\nu\varphi) +C_3R(\Box\varphi)
$$
\beq
+ B_1(\Box\varphi)^2+B_2(\nabla_\mu\varphi)^2(\Box\varphi)+
B_3(\nabla_\mu\varphi)^4+B_4m^2(\nabla_\mu\varphi)^2+B_5m^4],
\label{z11}
\eeq
where
$$
A_1={1\over 90}-{1\over{2b_1}}a''_1
$$
$$
A_2=-{1\over 90}-{1\over{2b_1}}a''_2+{1\over 24}({c_2\over b_1})^2+{c_2\over
{6b_1}}
$$
$$
A_3={1\over 36}-{1\over{2b_1}}a''_3+{1\over{48b^2_1}}[(4c'_3-4c_1-c_2)^2+
4(c'_3-c_1)(2c'_3-2c_1-c_2)]-{1\over{6b_1}}(c_1-c'_3+{1\over 2}c_2)\;
$$
$$
A_4=-{1\over{2b_1}}a''_4-{1\over{4b^2_1}}b_4(4c'_3-4c_1-c_2)-{1\over{6b_1}}
b_4
$$
$$
B_1=-{b''_1\over{2b_1}}+{1\over{4b^2_1}}(8(b'_1)^2-10b'_1b_2+5b^2_2)
$$
$$
B_2=-\frac{b''_2}{2b_1}+\frac{1}{2b^2_1}(b''_1b_2+4b'_1b'_2-b_2b'_2
-10b'_1b_3+10b_2b_3)-\frac{1}{2b^3_1}(2(b'_1)^2b_2+b^2_2b'_1)
$$
$$
B_3=-\frac{b''_3}{2b_1}+\frac{1}{3b^2_1}(5b''_1b_3+{3\over 4}(b'_2)^2
-5b'_2b_3+15b^2_3+5b'_1b'_3+b_2b'_3)$$ $$-\frac{1}{6b^3_1}(20(b'_1)^2b_3+
4b'_1b_2b_3+b''_1b^2_2+2b'_1b'_2b_2)+\frac{b^2_2(b'_1)^2}{2b^4_1}
$$
$$
B_4=-\frac{b''_4}{2b_1}+\frac{1}{2b^2_1}(4b''_1b_4-4b'_2b_4+6b_3b_4-5b'_1
b'_4-3b'_4b_2)+\frac{1}{b^3_1}(3b'_1b_2b_4-5(b'_1)^2b_4)
$$
$$
B_5=-\frac{b''_5}{2b_1}+{1\over 2}(\frac{b_4}{b_1})^2
$$
$$
C_1=-\frac{c''_1}{2b_1}-\frac{2b_3}{3b_1}+\frac{1}{b^2_1}
({1\over 2}b'_2c'_3
-3c'_3b_3+{1\over 2}b''_1c_1-{1\over 2}c_1b'_2+3c_1b_3+
{1\over 6}b''_1c_2
-{1\over 6}c_2b'_2+{2\over 3}c_2b_3
$$
$$
+c'_1b'_1+{1\over 6}b'_1b_2-{1\over 12}c'_2b_2+{1\over 12}c'_2b'_1)
+\frac{1}{6b^3_1}(b'_1b_2c_2-2(b'_1)^2c_2-6(b'_1)^2c_1)
$$
$$
C_2=-\frac{c''_2}{2b_1}+\frac{2b_3}{3b_1}+\frac{1}{6b^2_1}(5b'_1c'_2+
2c_2b_3+c'_2b_2-b^2_2)-\frac{b'_1b_2c_2}{3b^3_1}
$$
\beq
C_3=-\frac{c''_3}{2b_1}+\frac{1}{3b_1}(b'_1-b_2)+\frac{1}{6b^2_1}
(12b'_1c'_3-9b_2c'_3-9b'_1c_1+9c_1b_2-2b'_1c_2+2c_2b_2). \label{z12}
\eeq

Let us now  briefly analyze the above expression. First of all, notice
that  the
divergences (\ref{z11}), (\ref{z12}) have just the same general
structure as the classical
action (\ref{1x}). This fact indicates that the theory under consideration is
renormalizable, what is in  full accord with the more direct analysis
based on power counting. All the divergences can be removed by a
renormalization transformation of the functions $a(\varphi),b(\varphi),
c(\varphi)$, in analogy with two-dimensional sigma models. We do not
include the renormalization of the quantum field $\varphi$, since in the
case of arbitrary $b_1$ it leads to unavoidable difficulties.
Let us now say some words about the possible role of the matter fields.
Suppose that the dilaton model under consideration is coupled to a set
of free massless matter fields of spin $0, \frac{1}{2},1$. Then the
matter fields
contributions to the divergences of vacuum type lead to the following
change of the functions $A_{2,3}(\varphi)$ (see, for example, \cite{11}).
\bea
A_2&\rightarrow& A_2 = A_2+\frac{1}{60}\left( N_0+6N_{1/2}+12N_1
\right), \nn \\
A_3&\rightarrow& A_3 = A_3-\frac{1}{180}\left( N_0+6N_{1/2}+12N_1
\right) + \frac{1}{2} \left( \xi - \frac{1}{6} \right)^2 N_0,
\label{new1}
\eea
where $ N_0,N_{1/2}$ and $12N_1$ are the numbers of fields with the
corresponding spin,
and $\xi$ is the parameter of the non-minimal interaction in the scalar
field
sector. Here we have omitted the topological Gauss-Bonnet
 term for simplicity.
Thus, we see that even in the presence of the matter fields all the
divergences
can be removed by the renormalization of the functions $a(\varphi),b(\varphi),
c(\varphi)$. (In the case massive scalars and spinors a matter
contribution to $B_5$ and $A_4$ will also appear). Below it will be
shown that the above change
of $A_{1,2,3}(\varphi)$ does {\it not} affect our results seriously.

It is important to notice that
renormalization of the generalized couplings $a(\varphi),b(\varphi),
c(\varphi)$ explicitly manifests the properties which are usual for any
quantum field
theory in an external gravitational field \cite{3}. All these functions can be
easily separated into three groups, with a different renormalization
rule.
The first group is constituted by the $b(\varphi)$ functions. The
renormalization
of these functions is independent of the other functions, $a(\varphi),
c(\varphi)$, and is similar to the renormalization of matter fields couplings
in  usual models (like the $\phi^4$ coupling constant in the case of an
ordinary scalar field). The second group are the $c(\varphi)$
functions, which renormalize in a manner similar to that for the
nonminimal
constant $\xi$ of the  $\xi R\phi^2$ interaction \cite{3}. This
means that their
renormalization transformations are independent on $a(\varphi)$, but strongly
depend on $b(\varphi)$. The third group of
couplings is composed by the $a(\varphi)$, and they are similar to the
parameters of the action of the vacuum for ordinary matter fields.
Furthermore, the renormalization of the dimensionless functions
does not
depend on that of the dimensional ones, $a_4, b_4, b_5$, what is in
good accord with a well-known general theorem \cite{tyu}.
Thus, the theory under consideration possesses all the standard
properties of
the models on a curved classical background. The only distinctive
feature of the present one is
that the couplings in our theory are arbitrary functions of the field
$\varphi$. This fact can be interpreted as pointing out to the presence
of an infinite number of coupling constants.

Since the theory is renormalizable, one can formulate the
renormalization
group equations for the effective action and couplings and then explore
its asymptotic behaviour. The renormalization
group equations for the effective action have the standard form, since
the
number (finite or infinite) of coupling constants is not essential for the
corresponding formalism \cite{3}. The general solution of this equation
has the form
\beq
\Gamma[e^{-2t}g_{\alpha\beta},a_i,b_j,c_k,\mu] =
\Gamma[g_{\alpha\beta},a_i(t),b_j(t),c_k(t),\mu ],
\label{z112}
\eeq
where $\mu$ is the renormalization parameter and the
effective couplings satisfy renormalization group equations
of the form
$$
{da_i(t)\over{dt}} = \beta_{a_i},\;\;\;a_i=a_i(0),
$$
$$
{db_i(t)\over{dt}} = \beta_{b_i},\;\;\;b_i=b_i(0),
$$
\beq
{dc_i(t)\over{dt}} = \beta_{c_i},\;\;\;\;\;\;\;c_i=c_i(0).
\label{z113}
\eeq
Note that we do not take into account the dimensions of the functions
$a_4, b_4, b_5$. In fact we consider here these quantities as dimensionless
and suppose that the dimension of the corresponding terms in the action is
provided by some fundamental nonrenormalizable constant.
The beta-functions are defined in the usual manner. For instance,
\beq
\beta_{b_1}=\lim_{n\rightarrow 4} \mu \frac{d b_1}{d\mu}.
\label{z114}
\eeq
The derivation of the $\beta$-functions is pretty the same as in theories
with finite number of couplings, and we easily get
\beq
\beta_{a_i} = - (4\pi)^{-2} A_i,\;\;\;
\beta_{b_i} = - (4\pi)^{-2}B_i,\;\;\;
\beta_{c_i} = - (4\pi)^{-2}C_i.
\label{z115}
\eeq
In the next sections we shall present the analysis of the
renormalization group equations (\ref{z113}),(\ref{z115}).
In accordance with the considerations above, one can first explore the
equations for the effective couplings $b_{1,..,5}$, then for
$c_{1,2,3}$ and finally for the ``vacuum" ones $a_{1,..,4}$.
All that analysis looks much more simple for the conformal version of
the theory.

\medskip

\section{The conformally-invariant theory and some explicit solutions}

Let us now consider the most general conformally-invariant
version of the theory (\ref{1x}):
\beq
S_c = \int d^4x \, \sqrt{-g} \, \left\{ f (\varphi ) \varphi \nabla^4
\varphi  +  q (\varphi ) C^2_{\mu\nu\alpha\beta}+ p(\varphi)\left[
 \left( \nabla_\mu \varphi
\right) \left( \nabla^\mu \varphi \right) \right]^2 \right\}.
\label{21}
\eeq
Here $ f(\varphi ), q( \varphi)$ and $ p( \varphi)$
are arbitrary functions, $\nabla^4 = \Box^2 +2R^{\mu\nu} \nabla_\mu
\nabla_\nu - \frac{2}{3} R \Box + \frac{1}{3} (\nabla^\mu R)
\nabla_\mu$ is a fourth-order conformally invariant operator, and we
should recall that due to the fact that $[\varphi ]=0$ the conformal
transformation of our dilaton is trivial:
\beq
g_{\mu\nu} \longrightarrow e^{-2 \sigma}g_{\mu\nu}, \ \ \  \ \varphi
 \longrightarrow \varphi.
\label{22}
\eeq
Now, using expressions (\ref{2x}), one can integrate by parts the rhs
in (\ref{21}) and
present the result as a particular case of the theory (\ref{1x}), with
$$
a_1(\varphi) = q(\varphi),\;\;\;a_2(\varphi) = -2q(\varphi),\;\;\;
a_3(\varphi) = \frac{1}{3}q(\varphi),\;\;\;
$$
$$
b_1(\varphi) = f'(\varphi) \varphi +f(\varphi),\;\;\ b_2(\varphi) =
f''(\varphi) \varphi +2f'(\varphi) = b_1'(\varphi),\;\;\ b_3(\varphi) =
p(\varphi), $$
\beq
c_1(\varphi) = \frac{2}{3}f'(\varphi) \varphi +
\frac{2}{3}f(\varphi)=-\frac{1}{3} c_2,\;\;\;
c_2(\varphi) =-2 f'(\varphi) \varphi - 2f(\varphi).
 \label{23}
\eeq
The rest of the generalized couplings $a_4,b_4,b_5,c_3$
are equal to zero.
So, the general action (\ref{1x}) is invariant
under the conformal transformation (\ref{22}) when the functions
$a_i, b_j, c_k$ obey the constraints (\ref{23}).

Substituting the relations (\ref{23}) into the general expression for
the divergences of the effective action, we get
the divergences of the conformal theory in the form (\ref{z11}),
where instead of (\ref{z12}) we have
$$
B_1=-{b''_1\over{2b_1}}+{3(b'_1)^2 \over{4b^2_1}},\;\;\;B_2=B'_1,
\;\;\;C_1=\frac{2}{3}B_1,\;\;\;C_2=-2B_1,\;\;\;B_4=B_5=0,
$$
$$
B_3=-\frac{b''_3}{2b_1}+\frac{1}{3b^2_1}\left[ {3\over 4}(b''_1)^2
+15b^2_3+6b'_1b'_3\right] -\frac{1}{2b^3_1}\left[ 8(b'_1)^2b_3+
b''_1(b'_1)^2\right] +\frac{(b'_1)^4}{2b^4_1},
$$
\beq
A_1={1\over 90}-{1\over{2b_1}}a''_1,\;\;\;A_2=-2A_1,\;\;\;A_3=\frac{1}{3}A_1,
\;\;\;A_4=A_5=0.                   \label{100}
\eeq

As one can see from the last expressions,
the divergences of the
conformally-invariant theory (\ref{21}) apeear also in a conformally
invariant form  (up to the total divergence), as it should be.
The functions $A(\varphi),B(\varphi),
C(\varphi)$ obey the same conformal constraints (\ref{23}) with some
$F(\varphi),Q(\varphi),P(\varphi)$ instead of $f(\varphi),q(\varphi),
p(\varphi)$. Below, we shall use the functions $a_i, b_j, c_k$, taking
into
account the restrictions (\ref{23}), because in this way
calculations become more compact.
Hence, we have shown that the conformal invariant, higher-derivative
scalar theory considered here is renormalizable at the one-loop level
in a conformally invariant way, and therefore it is multiplicatively
renormalizable at one-loop. One can suppose that the general proof of
one-loop conformal renormalizability in an external gravitational
field, given in \cite{18} (see also \cite{3}), is valid for the
higher-derivative dimensionless scalar field as well.
Note that taking into account the matter field cotributions
does not lead, according to (\ref{new1}), to the violation of conformal
invariance. Indeed the conformal value of $\xi = \frac{1}{6}$ must be
choosen.

{}From a technical point of view, the
cancellation of non-conformal divergences gives a very effective tool
for the verification of the calculations. It moreover enables us to
hope that the general dilaton model (\ref{1x}) might be asymptotically
conformal invariant \cite{26,3}, just as the special case considered in
\cite{6}.

As a byproduct, the above expression also gives us the conformal
 anomaly of
the conformal invariant theory (\ref{21}): $T_\mu^\mu$ is equal to the
integrand of (\ref{z11}), (\ref{100})
(up to total derivatives, that we have dropped).
Thus, if one finds the form of the functions $f(\varphi),q(\varphi),
p(\varphi)$ which provide the one-loop finiteness in the theory
(\ref{21}), the last will be free from the conformal anomaly. Actually,
the one-loop effective action obeys the equation
\beq
-{2\over\sqrt{-g}}
g_{\mu\nu}{\delta\Gamma^{(1)}\over{\delta g_{\mu\nu}}}=T^{(1)},
\label{101} \eeq
where $T^{(1)}$ is the one-loop part of the anomaly trace of the
energy-momentum tensor. Eq.
(\ref{101}) allows one to define $\Gamma^{(1)}$ with accuracy up
to some conformally invariant  functional. Hence if we find the solution
of the equations
\beq
A_i \left( f(\varphi),q(\varphi),p(\varphi) \right) =
B_j \left( f(\varphi),q(\varphi),p(\varphi) \right)=
C_k \left( f(\varphi),q(\varphi),p(\varphi) \right)=0,
\label{102} \eeq
taking into account the constraints (\ref{23}), the
right-hand side of Eq. (\ref{101}) will be zero (up to surface terms),
and
$\Gamma^{(1)}$ will be a conformally invariant (but probably nonlocal)
functional. So the solution
gives us the conformal invariant theory (\ref{21}) that is free from the
anomaly (at least on the one-loop level). Moreover, according to the
structure
of the conformal Ward identities \cite{18}, \cite{3} it is clear, that
the two-loop divergences of the corresponding theory will be
conformally invariant as well.

The conditions (\ref{102}) are nothing but a set of nonlinear (and
rather complicated)
ordinary differential equations. Fortunately, one can use the results of
the qualitative analysis of the previous section and divide the equations
into three groups $B_j=0, C_k=0$ and $A_i=0$, respectively. It turns out
that
the only nontrivial problem is to explore the equations of the first group.
Note that, due to the conformal constraints (\ref{23}), the equation for
$b_3(\varphi)$ can be factorized out and we just have to deal with the
ones for
$b_1(\varphi)$ and $b_2(\varphi)$  first. Since the variable $b_2(\varphi)$
is not independent, we end up with only one equation for $b_1(\varphi)$
that
can be solved, in principle (actually just a very reduced number of
explicit solutions could be obtained, see below).
The only three solutions of power-like form are the following:
\beq
b_1 =k, \;\;\;\;\;k=\mbox{const.} ,\;\;\;\;\; \;b_2 = 0, \;\;\;\;\;
 b_3= 0,
\label{fcs1}
\eeq
\beq
b_1 =k, \ \ \ \ \ \ b_2 ={b_1}'= 0, \ \ \ \ \  \  b_3= \frac{3k}{5
 (\varphi - \varphi_0)^2},\;\;\;\;\; \varphi_0=\mbox{const.},
\label{cfp1}
\eeq
and
\beq
b_1 = \frac{k^2}{ (\varphi - \varphi_0)^2}, \ \  \ \ b_2 ={b_1}'
 =- \frac{2 k^2}{ (\varphi - \varphi_0)^3} , \ \ \  \  b_3=
 \frac{k^2}{ (\varphi - \varphi_0)^4}.
\label{cfp2}
\eeq
We should observe that  the second solution
(\ref{cfp1}) is a particular point of a whole surface of conformal fixed
points (i.e., conformal solutions) which can be expressed as
\beq
b_1 =k, \ \ \ \ \ \ b_2 = 0, \ \ \ \ \  \  b_3= F^{-1} (\varphi -
\varphi_0),
\label{fcs}
\eeq
where the function $F(p) = x$ is the solution of the differential
equation $p''- \frac{10}{k} p^2=0$, and is given by the quadrature:
\beq
\pm \int \frac{dp}{\sqrt{ \frac{20}{3k}p^3 +c_1}} = x, \label{quad1}
\eeq
with $c_1$  an arbitrary constant.

Since within the conformal theory the functions $c_{1,2,3}$ are not
independent,
the corresponding equations are satisfied automatically. The equations for
$a_{1,2,3}$ have the following corresponding solutions. For (\ref{fcs1})
and (\ref{cfp1}), the common one
\beq
a_1(\varphi)=\frac{(\varphi)^2}{90}+a_{11}\varphi+a_{12},   \label{103}
\eeq
and for (\ref{cfp2}),
\beq
a_1(\varphi)=-\frac{1}{45}\ln|\varphi - \varphi_0|
+a_{11}\varphi+a_{12},   \label{104}
\eeq
where $a_{11}$ and $a_{12}$ are integration constants and
$a_{2}(\varphi)$ and $a_{3}(\varphi)$ are both defined via the conformal
constraints (\ref{23}). Let us notice that the above finite solutions
(with evident numerical modifications) is stable
under the contributions of the matter fields, that directly follows
from (\ref{new1}).

In this way we have constructed three explicit examples of one-loop
finite,
anomaly free, conformal theories. The model (\ref{fcs1})  is
essentially the
same theory which had been investigated in
previous articles \cite{6}. It is closely related with the theory of
induced conformal factor \cite{antmot,rei,frts,27}. Since the only
nontrivial interactions here are of ``nonminimal" and ``vacuum" type,
it is renormalized in a manner similar to the one for
the theory of a free (ordinary)
scalar field in an external metric field. That is why the
finiteness of this model is rather trivial.
Not so are the solutions (\ref{cfp1}), (\ref{cfp2}) and (\ref{fcs}).
These models contain
nontrivial interaction sectors and their finiteness does not look trivial
at all.
Moreover, the form of solution (\ref{cfp2}) probably indicates
that some extra
symmetry is present. Notice also that both nontrivial solutions
depend on
the arbitrary value $\varphi_0$ and are singular in the vicinity of this
value. One could argue that this fact hints towards the existence of
some different, nonsingular parametrization of the field variable.
This conformally invariant finite model might be quite interesting in
connection with some attempts to generalize the $C$-theorem \cite{29} to
four dimensions \cite{30}.

\medskip

\section{Explicit non-conformal solutions}

We now turn to the search of finite solutions of the
general model (\ref{1x}), (\ref{z12}), free of the conformal
constraints.
Since we are looking for non-conformally invariant solutions,
${b_1}'$ must be different from $b_2$ (otherwise we
get back to the conformally invariant case).
{}From the mathematical point of view, to obtain solutions of the
general system (\ref{102}), (\ref{z12})
is a rather difficult problem, because in this case
the equation $B_3=0$ is not factorized out. So, already at a first
stage, we are
faced up with a set of the nonlinear, higher-dimensional differential
equations.
Fortunately, these equations exhibit some homogeneity property, and
hence it is natural to look for  solutions of the exponential form
$b_j(\varphi)=k_j \exp \left[ (\varphi - \varphi_0)\lambda_j \right]$,
and of the power-like form
$b_j(\varphi)=k_j   (\varphi - \varphi_0)^{\lambda_j} $,
where $k_j$ and $\lambda_j$ are some constants.

Accurate analysis shows that all the $\lambda_j$ are necessary equal to
zero in the exponential case. Quite on the contrary, the search for
solutions
of power type yields the following three non-conformal fixed points:
\beq
b_1 = \frac{k}{ (\varphi - \varphi_0)^{5/3}}, \ \  \ \ b_2
 =- \frac{2 k}{ (\varphi - \varphi_0)^{8/3}} , \ \ \  \  b_3=
 \frac{16k}{15 (\varphi - \varphi_0)^{11/3}},
\label{ncfp1}
\eeq
\beq
b_1 = \frac{k}{ (\varphi - \varphi_0)^{5/3}}, \ \  \ \ b_2
 =- \frac{4 k}{3 (\varphi - \varphi_0)^{8/3}} , \ \ \  \  b_3=
 \frac{4k}{9 (\varphi - \varphi_0)^{11/3}},
\label{ncfp2}
\eeq
and
\beq
b_1 = \frac{k}{ (\varphi - \varphi_0)^{1/3}}, \ \  \ \ b_2
 =0 , \ \ \  \  b_3=0.
\label{ncfp3}
\eeq

These are in fact the {\it only} solutions of power type.
The solution of the equations for $a_i(\varphi)$ and $c_k(\varphi)$
is then straightforward (but involved). We shall present only the
results of this analysis.
For all three solutions (\ref{ncfp1})--(\ref{ncfp3}), the $a_i(\varphi)$
are given by the integrals
\beq
a_i=\int\limits_{\varphi_{i1}}^{\varphi}\int\limits_{\varphi_{i2}}^{\varphi}
2b_1(\varphi)\left[A_i(\varphi)+\frac{1}{2b_1(\varphi)}a''_i(\varphi)
\right].
\label{ncfp7}
\eeq
Notice that in the last expression the integrands do not depend on
$a_i(\varphi)$ while
$\varphi_{i1}$ and $\varphi_{i2}$ are arbitrary constants.
In the case of the theory coupled to matter fields the values of
$A_{1,2,3}$ have to be substituted according to (\ref{new1}). The
solutions for $b_{4,5}$ and $ c_{1,2,3}$ are written below.
For the case (\ref{ncfp3}), these solutions have the form
$$
c_2\;=\;r_0 x^{\frac{4}{9}}+r_1
$$
$$
c_1\;=\;-\frac{2}{7}r_0 x^{\frac{4}{9}}-\frac{1}{3}r_1
+r_2 x^{\frac{2}{3}}+r_3x^{-\frac{1}{3}}
$$
$$
c_3\;=\;-\frac{45}{364}r_0 x^{\frac{13}{9}}-\frac{1}{12}r_1x
-\frac{9}{10}r_2 x^{\frac{5}{3}}+\frac{2k-9r_3}{10}x^{\frac{2}{3}}
+r_4+r_5 x^{\frac{7}{3}}
$$
$$
b_4\;=\;r_6 x^{\frac{4+\sqrt{22}}{3}}+r_7 x^{\frac{4-\sqrt{22}}{3}}
$$
$$
b_5\;=\;r_8+r_9x+\frac{9}{44}r_6 r_7 x^{\frac{14}{3}}
$$
\beq
+\frac{r_6^2}{k\left( 4+\frac{2}{3}\sqrt{22} \right)
\left( 5+\frac{2}{3}\sqrt{22} \right)} x^{5+\frac{2}{3}\sqrt{22}}+
\frac{r_7^2}{k\left( 4-\frac{2}{3}\sqrt{22} \right)
\left( 5-\frac{2}{3}\sqrt{22} \right)} x^{5-\frac{2}{3}\sqrt{22}},
\label{ncfp4}
\eeq
where, for the sake of brevity, we have denoted $x\equiv\varphi -
\varphi_0$
 and introduced the set of integration constants $r_0,...,r_9$.
For the cases when the $b_i$ are given by (\ref{ncfp1}) and
(\ref{ncfp2}), the solutions for $c_i$, $b_4$ and $b_5$ are still
easily found in a closed form, but we will not bother the reader with
such lengthy espressions here.
Thus we have constructed the finite nonconformal versions of the theory
(\ref{1x}). The functions $a_i(\varphi)$,
$b_j(\varphi)$ and $c_k(\varphi)$
 above   correspond to the finite theory.

\medskip

\section{Renormalization group and stability analysis}

Here we apply a method of analysis based on the renormalization group
for the investigation of the general model (\ref{1x}). If we do not impose
the conditions (\ref{102})  on the interaction functions,
then the theory is not finite (of course, it is possible that
there exist some other nonconformal finite solutions) but
renormalizable.
As it was already pointed out above, the renormalization group
$\beta$-functions
are defined in a unique way (\ref{z115}), and we arrive at the following
renormalization group equations for $a_i(\varphi)$,
$b_j(\varphi)$ and $c_k(\varphi)$:
\beq
\frac{da_i}{dt'}\;=\;- A_i,\;\;\;\;\;
\frac{db_j}{dt'}\;=\;- B_j,\;\;\;\;\;
\frac{dc_i}{dt'}\;=\;- C_i,
\label{z116}
\eeq
where $t'=(4\pi)^{-2}t$, and $t$ is the parameter of the rescaling of
the background metric (\ref{z12}).

The renormalization group equations (\ref{z116}) have a complicated
structure. In fact the effective couplings $a,b,c$ depend not only on $t$,
but also on $\varphi$ and, therefore, (\ref{z116}) is nothing but a
set of nonlinear, higher-order differential equations in terms of
partial derivatives.
For this reason, to obtain the complete solution of these equations does
not seem to be possible.
At the same time, we already know the values of $a,b,c$ which correspond to
vanishing $\beta$-functions. From the renormalization group point
of view these values are the fixed points of the theory. Thus, we
can explore the stability of the fixed points (\ref{z116})
and then  formulate
some conjectures concerning the asymptotic behaviour of the theory.

We thus face the problem of the stability analysis of a system with
an infinite
number of variables. A possible way to attack it consists in combining
the standard Lyapunov method and harmonic Fourier analysis. Let us
first
illustrate the method on the most simple example  of the conformal fixed
point (\ref{cfp1}). The advantage of this solution is that the equations
for $b_1$ and $b_3$ do not depend on each other. One can start with the
equation for $b_1$, and put $k=1$ for the sake of simplicity. Moreover,
we shall write $t$ instead of $t'$. According to the
Lyapunov method we write $b_1=1+y(x)$, where $x=\varphi-\varphi_0$ and
$y$ is the infinitesimal variation of $b_1$. Hence we preserve the
conformal
constraint $b_2=1+y'(x)$, where the derivative is taken with respect to
$x$. Substituting the above expressions into the renormalization group
equation, we get
\beq
\frac{dy}{dt}\;=\;\frac{1}{4(1+y)}
\left[ 2y''_{xx}(1+y)-\frac{3}{4}(y'_{x})^2 \right].  \label{z117}
\eeq
Since we are only interested in the behaviour at the vicinity of
the fixed
point, the nonlinear terms of the last equation can be safely omitted,
and we obtain
\beq
\frac{dy}{dt}\;=\;\frac{1}{2} y''_{xx}. \label{z118}
\eeq
This equation looks very simple but it depends still on two variables.
However (\ref{z118}) can be easily reduced to a set of ordinary
differential equations.
One can expand $y(x)$ in Fourier series with  $t$ dependent
coefficients:
\beq
y(x,t)\;=\;\frac{y_0(t)}{2}\;+\;\sum_{n=1}^{\infty}
y_n(t)\cos nx\;+\;{\wt{y}}_n(t)\sin nx.  \label{z119}
\eeq
Substituting (\ref{z119}) into (\ref{z118}) we obtain
\beq
\frac{dy_0}{dt}=0, \;\;\;\;\;\;\frac{dy_n}{dt}=-\frac{n^2}{2}y_n,
\;\;\;\;\;\; \frac{d\wt{y}_n}{dt}=-\frac{n^2}{2}{\wt{y}}_n.
 \label{z120}
\eeq
{}From (\ref{z120}) it follows
that all the coefficients except for $y_0$  vanish in the
limit $t\rightarrow+\infty$. Since the infinitesimal variation cannot
contain a zero mode, one can put $y_0 =0$ and hence the fixed value
$b_1=1$
is stable in the mentioned limit. Notice that if we do not input the
conformal
constraints, that is, if we take $\delta b_2 \neq (\delta b_1)'$, then
the values  $b_1=1,\;b_2=0$ give a saddle point of the theory.

Then one can start with $b_3$, what is a bit more complicated.
If one introduces the infinitesimal variation $z$ as
$b_3=\frac{5}{3(\varphi-
\varphi_0)}+z$ and omits all the nonlinear terms, the remaining equation is
\beq
\frac{dz}{dt}=\frac{1}{2}z''_{\varphi\varphi}-
\frac{50}{3} \frac{z^2}{\varphi-\varphi_0}.
 \label{z121}
\eeq
If we consider the behaviour of $z$ in a region far from the value of
$\varphi_0$,
then the factor $(\varphi-\varphi_0)^{-1}$ is slowly varying and one can
regard it as a constant $x_0$. After expanding $z$ into a Fourier series,
we get \beq
\frac{dz_0}{dt}=-\frac{50}{3} \frac{z_0}{x_0^2},
\;\;\;\;\;\;
\frac{dz_n}{dt}=\left(-\frac{1}{2}n^2-\frac{50}{3 x_0^2}\right)z_n,
\;\;\;\;\;\;
\frac{d\wt{z}_n}{dt}= \left(-\frac{1}{2}n^2-\frac{50}{3 x_0^2}\right)
{\wt{z}}_n,  \label{z122}
\eeq
what reveals the stable nature of this conformal fixed point.
The exploration of the behaviour of the $c_{1,2,3}$ is not necessary,
because they are related with $b_1$ by the conformal constraints,
and thus their behaviour is completely determined.

If one takes the values of the infinitesimal corrections which violate
the conformal constraints then this fixed point is a saddle one.
The last claim is actually trivial, since we already knew this from the
behaviour of $b_2$.
Stability analysis
performed on the last of the non-conformal
solutions (\ref{ncfp3}) shows that it is a saddle
point of the non-conformally invariant theory. It is clear already from the
behaviour of $b_1, b_2, b_3$ and hence further investigation is not
necessary.

So we can see that among the fixed points of the theory there are
some which are completely stable in UV limit and others which are
partially stable, namely saddle points of the renormalization group
dynamics.
One can conjecture that the behaviour of the functions $a(\varphi),
b(\varphi),
c(\varphi)$ essentially depends on the choice of the initial data (with
respect
to the renormalization group parameter), which have to be postulated
at some given energy. In particular, it is natural to expect that
for some conformal models at high energies, asymptotic finiteness
manifestly appears.
 Simultaneously, there is a cancellation of the conformal anomaly in
this limit. In such way, the theory (\ref{1x}) predicts the existence of
renormalization group flows from arbitrary values of $a,b,c$ to the one
which provides finiteness and conformal invariance of the theory.

\medskip

\section{Discussion}

We have investigated the  renormalization group
behaviour of the general dilaton model (\ref{1x}) on the background of a
classical metric. The theory under consideration possesses interesting
nontrivial
features, as finite fixed points and plausible renormalization
group
flows between these points. This fact has important physical
applications,
if we make use of the hypothesis in \cite{antmot} and regard the dilaton
theory as
an approximation to some more fundamental theory of quantum gravity
(like the theory of strings)  at low energies.
The action of gravity, induced by string loop effects, has the
form of a series in the string loop parameter
$\alpha'$, and at second order it contains the terms
with fourth derivatives of the target space metric and the dilaton
\cite{mets}. Thus, within some accuracy, the
effective action of the string is a particular case of our dilaton
model (the well known arbitrariness in the second order
effective action for the string does not affect our speculations here).
This particular case is not a fixed point of our model (maybe only
at one loop). One can suppose that our theory of the dilaton
is valid at scales between the
Planck  energy $M_p$ and some energy $M_l$, where the effects of
quantum gravity
are weak and only matter fields can be regarded as quantum ones.
It is rather remarkable that the action for the dilaton ---generated by
quantum
effects of the matter fields--- is an IR fixed point of our general
dilaton model. Hence our dilaton model
can  describe the
transition from string induced dilaton gravity at the $M_p$ scale to
matter induced gravity at the $M_l$ scale.

We can also say some words about the expected effects of the quantum
metric.
In spite of the fact that the theory (\ref{1x}) is rather involved,
one can calculate
the one-loop divergences with the use of the method proposed in \cite{first}.
Moreover, some conjectures concerning the renormalization of the
theory of quantum gravity based on (\ref{1x}) can be made even without
 carring out calculations to the end explicitly.
As has been already pointed out above, the general structure of the
expressions
for the counterterms will be similar to (\ref{z12}). This means
that all
the structures (but not necessarily the
numerical coefficients,
of course) will be actually the same. However, the structure of the
renormalization might
be much more complicated. In particular, for the theory of quantum
gravity  the
 hierarchy of the couplings is lacking and all the functions $b,c,a$
have to be renormalized simultaneously, what is rather more cumbersome
 as compared with the dilaton theory  described above.
However, the general structure of the counterterms in the case of
the quantum metric must be the same as for our dilaton model.
In particular, the functions $A_i, B_j, C_k$ are expected to be
homogeneous just as in the case considered above.
Hence one can hope to get similar finite solutions in the general
theory.

The final point of our discussion
is related with the conformal invariance properties at the quantum level.
  Some features of the theory of quantum gravity  based on
(\ref{21})
(with $p(\varphi)=0$) have been recently discussed in \cite{17}.
It was shown there that, generally, the theory leads to a conformal
anomaly. This
anomaly appears already in the one loop counterterms and prevents the
theory from being renormalizable. For this reason, we cannot
expect from a theory of quantum gravity
based on (\ref{1x}) to have a conformal fixed point.
However, it should be possible to obtain
conformal invariance
at the quantum level within the general model (\ref{1x}), by introducing
the loop expansion parameter in an explicit way.

It would be of interest to study the cosmological consequences that
arise from the family of finite models (1), as the possible existence of
solutions of black hole type and their influence on the evolution of the
early universe.
The dilaton in the starting theory is massive, owing to the nontrivial
dimensions
of the functions $b_4(\phi), b_5(\phi)$. For the finite conformal versions
of the theory it is not more so. However one can get massive
parameters
as a result of some symmetry breaking and for this purposes it is
necessary, for instance,
 to derive the effective potential and to explore the possibility
of a phase transition (see \cite{coind,buchodsh} for the discussion of
that approach). Indeed, the effective potential in dilatonic gravity
under discussion has the form (in the linear curvature approximation)
\[
V=b_5(\varphi) + \frac{1}{2} B_5 (\varphi) \ln \frac{\chi
(\varphi)}{\mu^2} + R \left[ a_4 (\varphi) + \frac{1}{2} A_4(\varphi)
\ln \frac{\chi (\varphi)}{\mu^2} \right],
\]
where $\chi (\varphi)$ is some combination of the dimensional functions
$a_4$ and $b_4$. Its explicit form plays no role in this qualitative
discussion. It is clearly seen that the one-loop level effective action
of our theory at low energies represents the standard Einstein theory
with $\varphi$-dependent cosmological and gravitational constants.
Hence, our theory leads to the induction of general relativity at low
energies, what serves as an additional physical motivation for its
detailed study.
Notice also that one can introduce massive
terms even in the
conformal case, what is something like soft breaking of the
conformal invariance. It is possible to provide finiteness even in this
case
(as well as in nonconformal versions of the theory, of course).
We expect to return to such questions elsewhere.

Summing up, we have explored some features of the general dilaton
model (\ref{1x}) which can be regarded as a toy model for the same
theory with a quantum metric. Some special versions of the model are
finite at one loop and, moreover, some of them are conformally invariant
both at the classical and at the quantum level. The lack of conformal
anomaly
holds even if the matter field contributions are taken into account.
The last property is likely to
 survive for the more general model with a quantum metric. In this
respect the theory discussed above is the first example of such kind.
Furthermore we have investigated its stability of found several fixed
points
(developing by the way new mathematical tool for this purposes).
This enables
us to draw some conclusions on the possibility of renormalization
group flows between the different versions of the theory. In particular,
one can hope to apply our model to obtain the connection between the
string induced gravity action at the Planck energy scale and the matter
field
induced action at some lower scale, what is certainly valuable for
phenomenology purposes.

\vspace{5mm}

\noindent{\large \bf Acknowledgments}

 EE and ILS are grateful to
T. Muta  and to the whole Department of Physics, Hiroshima
University, for warm hospitality.
SDO would like to acknowledge the kind hospitality of the
members of the Department ECM, Barcelona University.
We thank also the referee for relevant comments to our previous version
of the paper. This work has been
supported  by the SEP Program, by DGICYT
(Spain),  by CIRIT (Generalitat de Catalunya),
by the RFFR (Russia), project
no. 94-02-03234, and by ISF (Russia), grant RI1000.

 \end{document}